\documentclass[conference]{IEEEtran}
\usepackage{subfigure}
\IEEEoverridecommandlockouts
\usepackage{cite}
\usepackage{amsmath,amssymb,amsfonts}
\usepackage{algorithm}
\usepackage{algorithmic}
\usepackage{graphicx}
\usepackage{textcomp}
\def\BibTeX{{\rm B\kern-.05em{\sc i\kern-.025em b}\kern-.08em
    T\kern-.1667em\lower.7ex\hbox{E}\kern-.125emX}}

\usepackage{mathptmx}
\DeclareGraphicsExtensions{.pdf,.png,.jpg}
\graphicspath{{.pic/}}
\usepackage{comment}
\usepackage{soul}
\usepackage[dvipsnames]{xcolor}
\usepackage{listings}
\newcommand{\id}[1]{\ensuremath{\mathit{#1}}}
\newcommand{\FIG}[1]{Fig.~\ref{#1}}

\specialcomment{cmtPai}{\begingroup\sffamily\scriptsize\color{red}}{\endgroup}
\specialcomment{cmtNew}{\begingroup\sffamily\scriptsize\color{OliveGreen}}{\endgroup}
\specialcomment{cmtAli}{\begingroup\sffamily\scriptsize\color{Green}}{\endgroup}

\begin{document}
\title{\huge{HEAD-MOUSE: A SIMPLE CURSOR CONTROLLER BASED ON OPTICAL MEASUREMENT OF HEAD TILT}}
\author{
\IEEEauthorblockN{Ali HeydariGorji$^{1}$, Seyede Mahya Safavi$^{1}$, Cheng-Ting Lee$^{2}$, Pai H. Chou$^{2}$}
\IEEEauthorblockA{$^{1}$ Department of Electrical Engineering and Computer Science, University of California, Irvine, USA\\
$^{2}$ Department of of Computer Science National Tsing Hua University, Hsinchu City, 30013 Taiwan\\
 e-mail: \textit{\{heydari, safavis\}}@uci.edu, \textit{\{davis0704.lee\}}@gmail.com, \textit{\{phchou\}}@cs.nthu.edu.tw}
}

\maketitle
\begin{abstract}

This paper describes a wearable wireless mouse-cursor controller that
optically tracks the degree of tilt of the user's head to move the mouse
relative distances and therefore the degrees of tilt.  The raw data can be
processed locally on the wearable device before wirelessly transmitting the
mouse-movement reports over Bluetooth Low Energy (BLE) protocol to the host
computer; but for exploration of algorithms, the raw data can also be
processed on the host.  The use of standard Human-Interface Device (HID)
profile enables plug-and-play of the proposed mouse device on modern computers
without requiring separate driver installation. It can be used in two
different modes to move the cursor, the joystick mode and the direct mapped
mode.  Experimental results show that this head-controlled mouse to be
intuitive and effective in operating the mouse cursor with fine-grained
control of the cursor even by untrained users.

\end{abstract}


\section{Introduction}
\label{sec:intro}

Pointing on a two-dimensional display has been a fundamental operation
in graphical user interface in the past three decades and will
continue to remain important in the foreseeable future.  Pointing
devices on a two-dimensional screen may have involved from desktop mouse and track balls to track
pads and touch screens. However, for patients suffering from amputation or paralysis,
alternative
mechanisms must be used to enable their operation with computers.

Assistive technologies developed for enabling hands-free
cursor control are mainly divided into eye trackers \cite{eye_tracking}, head trackers,
tongue trackers \cite{meysam}, brainwave (EEG) sensors \cite{EEG}, and muscle tension sensors
(EMG) \cite{EMG}. These techniques target users with a wide range of motor disabilities, and
they have all been shown to be effective in their own specific ways. However,
they all come at a relatively high cost. Camera-based
systems require continuous image processing, which requires nontrivial
computation power and can operate for at most hours on a battery. This makes them unsuitable for portable long-term usage in the range of weeks or months \cite{survey}. Tongue trackers may require special
sensors to be placed in the mouth and may be considered intrusive even
if not invasive.  In addition to a poor performance in terms of speed and
accuracy, EEG electrodes are extremely complex and uncomfortable due to the large number of electrodes and wires \cite{zoran}.  Although voice recognition
can handle text input for verbal commands and dictation with out any body
mounted extra device, it has a low degree of freedom. In addition, the computational load due to voice processing unit makes it power consuming and slow.

We envision a low-cost system that is comfortable to wear, intuitive to use, and require little or no
training or calibration.  To achieve these goals, we propose a simple
mechanism that requires minimal processing based on infra-red LED (IR LED) and photo detector pair. IR LED and photo detector pair can be an effective mechanism for precise short-range distance measurement.  The novelty with our
work is the adaptation of the active IR-based mechanism to measuring
distance from the collar area to the sides of the chin as our way of
tracking head tilt to control the mouse cursor.  The IR LED and photo
diodes are among the lowest-cost sensors, and the rest of the
functionality including the ADC, processor, and wireless communication
can all be done in commodity Bluetooth Low Energy (BLE)
microcontrollers on the order of \$2-3 in low-volume quantities.
Due to the simplicity of algorithms and calculations, they can be implemented on the MCU to eliminate the need for a third party software on the host system. As a result, the device can be directly connected to the host system using generic HID drivers.
It can also drives the cursor in two different modes for the ease of use. Both modes, the joystick and direct mapping, are explained in the further sections.
We have built a prototype of the proposed ``head mouse'' system and
tested it on a number of users.  Experimental results show that our
system to be intuitive to use, comfortable to wear, low power, and low
cost.

\section{System Design}
\label{sec:overview}

\FIG{system} shows the block diagram of the proposed system.  The wearable
system is centered around a microcontroller unit (MCU), sensors, a power source,
and a wireless communication interface.  The wearable system transmits either
raw or processed data to the host computer where the cursor is displayed.


\subsection{MCU and Wireless Communication}

We suggest an MCU-based embedded system.  On one side, it reads the signal from
the sensor for detecting the head position.  The MCU is programmable and can
process data.  On the other side, the MCU is connected to a wireless
communicating module to transfer the raw or processed data to the host system.
Many conventional MCUs can be used.  To be suitable for a wearable device,
we choose one with more integrated features, including on-chip analog-to-digital
converter (ADC) for interfacing with the optical sensors, and low power
consumption, rather than one with high performance.

The choice of wireless communication enables ease of use and mobility of device.
The device and the host need to support a compatible protocol.  We chose
Bluetooth Low Energy Technology (BLE), suitable for low-data-rate communication.
It consumes considerably less power compared to convectional Bluetooth modules.
Moreover, BLE is also integrated on a number of popular MCUs, making it low cost
and compact.

Our choice of MCU is the TI CC2541, although many other MCUs in the similar
class can also be used.  The processor is an 8051-compatible core surrounded by
8-KB SRAM, 256~KB flash, 8-channel ADC, timers, UARTs, I2C, and SPI on the
chip.  It also includes a BLE transceiver that enables connection to any
Bluetooth device including the users's computer. We used this SoC to transmit
data to the host system.

\subsection{Optical Sensor}

Our proposed system consists of two pairs of IR LED and photodiodes to estimate
the distance of the LED from the edges of user's chin.  The collected data is
digitized using an embedded analog-to-digital converter and sent through the BLE
module to the host computer. This subsection describes the sensors in detail.

\begin{figure}
\centering
  \includegraphics[bb=-568 -216 2046 693,trim=0 0 0 0,width=\columnwidth]{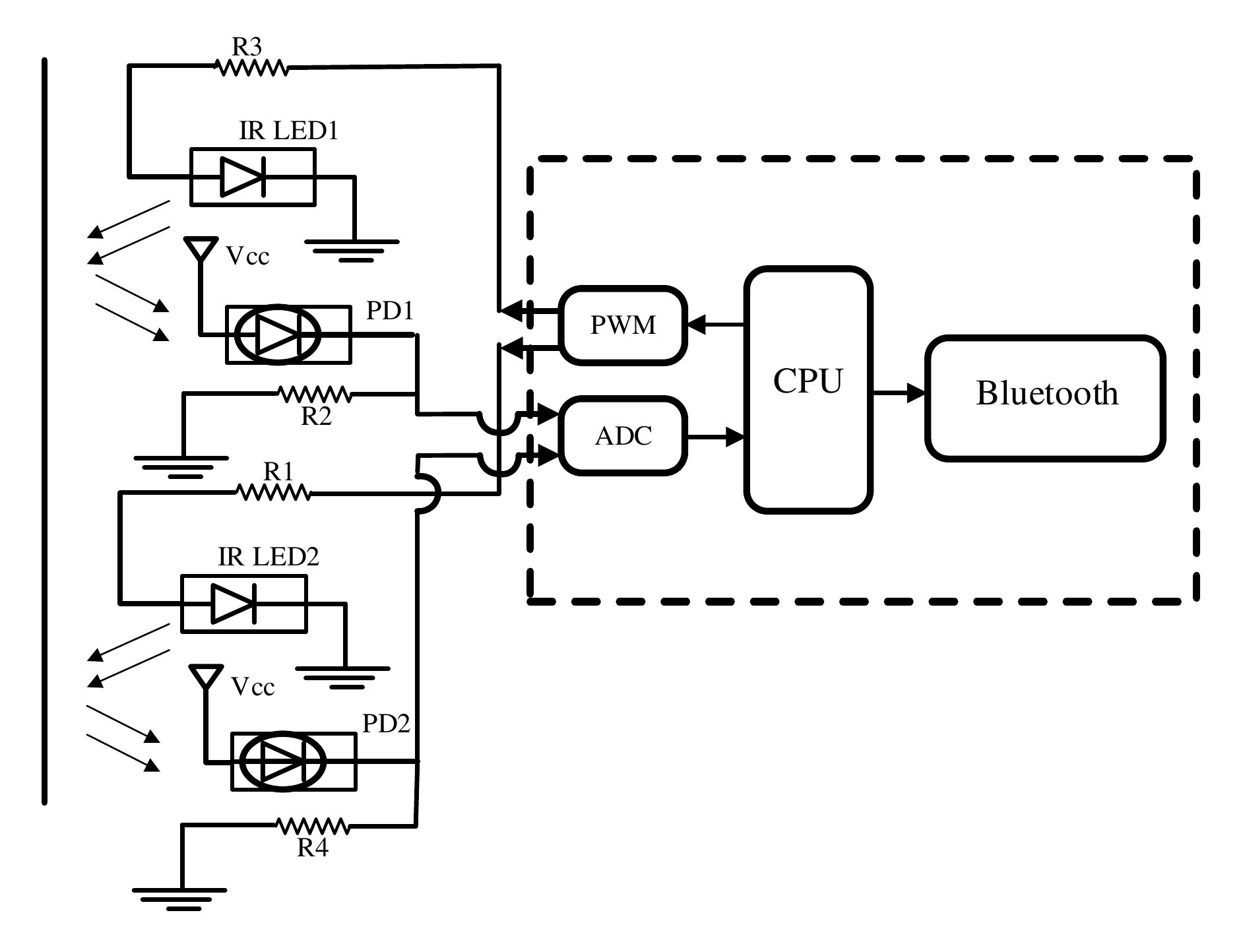}
  \caption{The system block diagram.}
  \label{system}
 \end{figure}

The IR transmitters and
receivers work on 940 nm wavelength and are available commercially
off-the-shelf. For simplicity, the transmitter is set
to be voltage-driven rather than current-driven using a 220~$\Omega$ shunt
resistor to control the current and IR brightness as shown in \FIG{system}.
The receiver is an IR photo-detector, modeled as a variable resistor. The
resistance is determined by the amount of IR that the photo-detector receives. A
shunt resistor of 22~K$\Omega$ is used to bias the output. The output voltage
changes approximately linearly with respect to the intensity of the IR light.
The IR intensity itself is a function of the distance between the sensor board
and the reflective surface (chin). Each sensor can detect the chin distance in
the range of 0.2 to 4 inches.

In order to make the device power efficient, pulse-width modulation (PWM) is
used.  Sensors are powered on once every 50~ms for a period of 1~ms; in other
words, 20~Hz sampling frequency and 5\% duty cycle.
This reduces power usage
considerably and increases battery endurance without significantly affecting
the
device's performance. Since the output signal is analog, it can have any value
in the range of VCC=3.3 volts and GND. The signal is converted using the on-chip
ADC on the MCU. The ADC samples each photo diode once every 50~ms at 8-bit
resolution using a time-multiplexing scheme.

\subsection{Power Subsystem}

The power subsystem consists of a battery and a power regulator.  We chose a
rechargeable lithium ion battery as the source of power.  The battery power
is fed to a buck converter, the TPS62740DSS, to reduce 3.8~V to 2.5~V to be
used by the MCU.  Note that IR transmitters are directly connected to 3.8~V
from the battery (and GPIO pins control the path to ground), rather than the
regulated 2.5~V, as the higher voltage results in more accurate distance
measurement.
  The battery can be charged by wired connectors using
off-the-shelf chargers. Wireless charging can be considered for future work.

\subsection{Mechanical Design}

\begin{figure}
\centering
  \includegraphics[trim=0 0 90 90,width=\columnwidth]{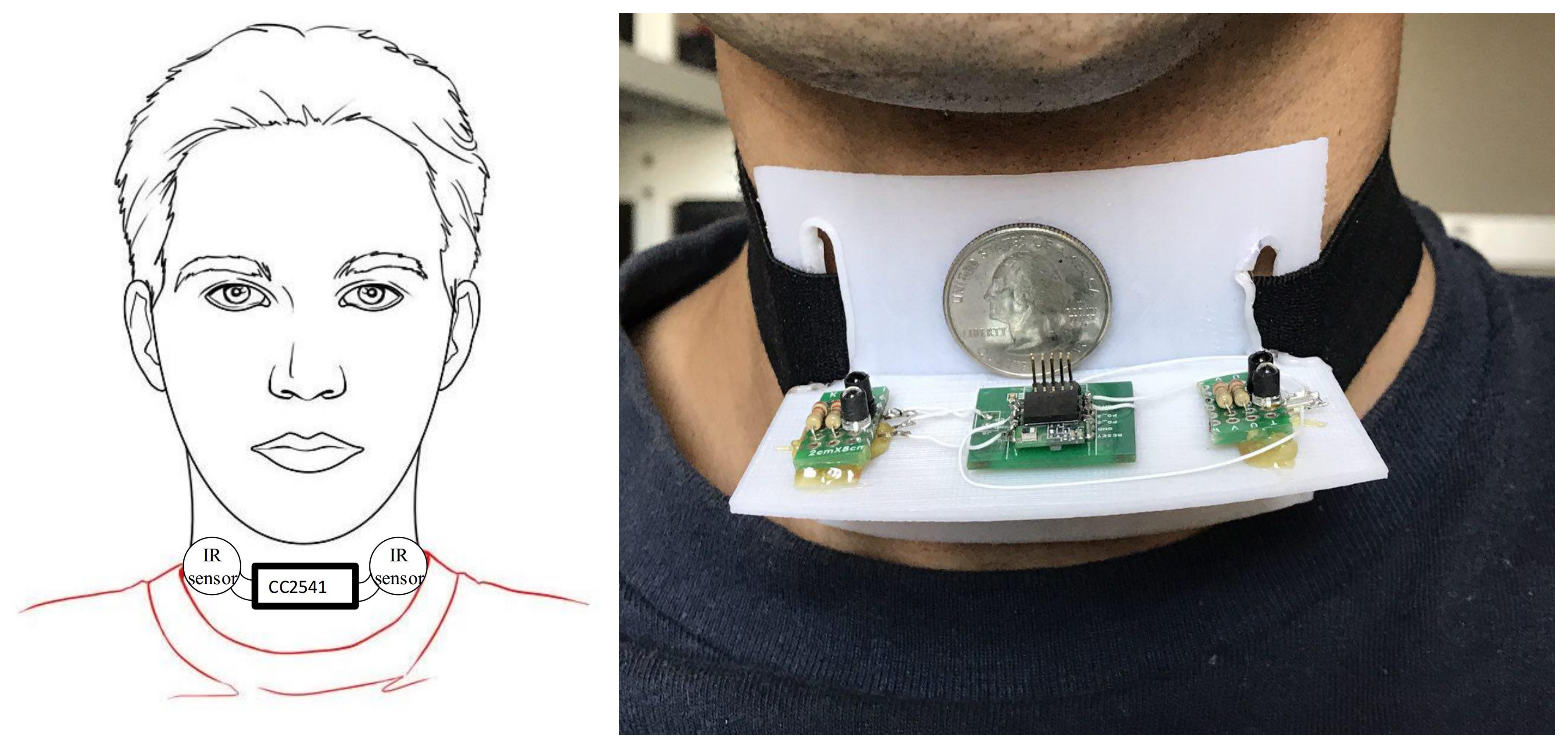}
  \caption{The location of the optical sensors and the CC2541 module
	including the BLE, ADC, and the battery.  A US quarter coin is placed for
size comparison but is not part of the system}
 \label{deck}
  \end{figure}

All of the components including the sensors, main board, and the battery are
placed on a curved-shaped deck that can easily be mounted on the neck as in
\FIG{deck}. The overall size of the device is $4\times7\times3$~cm$^3$. To
make the device compatible for different neck sizes, two fabric straps are
mounted on either side of the deck.

\subsection{Host Computer}

Depending on the mode of operation, further processing on the raw data may be
performed before transmission to the host. For now, raw data is being sent
directly and all the processing and algorithms are implemented on the host
system for easier exploration and better reproducibility, as we can run a
variety of algorithms on the same recorded data.   We have developed a
Python-based Graphical User Interface (GUI) for PC to receive data and draw
the correspond position on the screen.
We have also implemented the human interface device (HID) profile over BLE.
The HID profile enables our device to work as a wireless mouse over BLE and
enables plug-and-play operation without requiring driver
installation.  A later version will process the data on the device and send
mouse movement data to the host just as a mouse would.
    \begin{figure}
\centering
 \includegraphics[trim=70 50 40 50,width=\columnwidth]{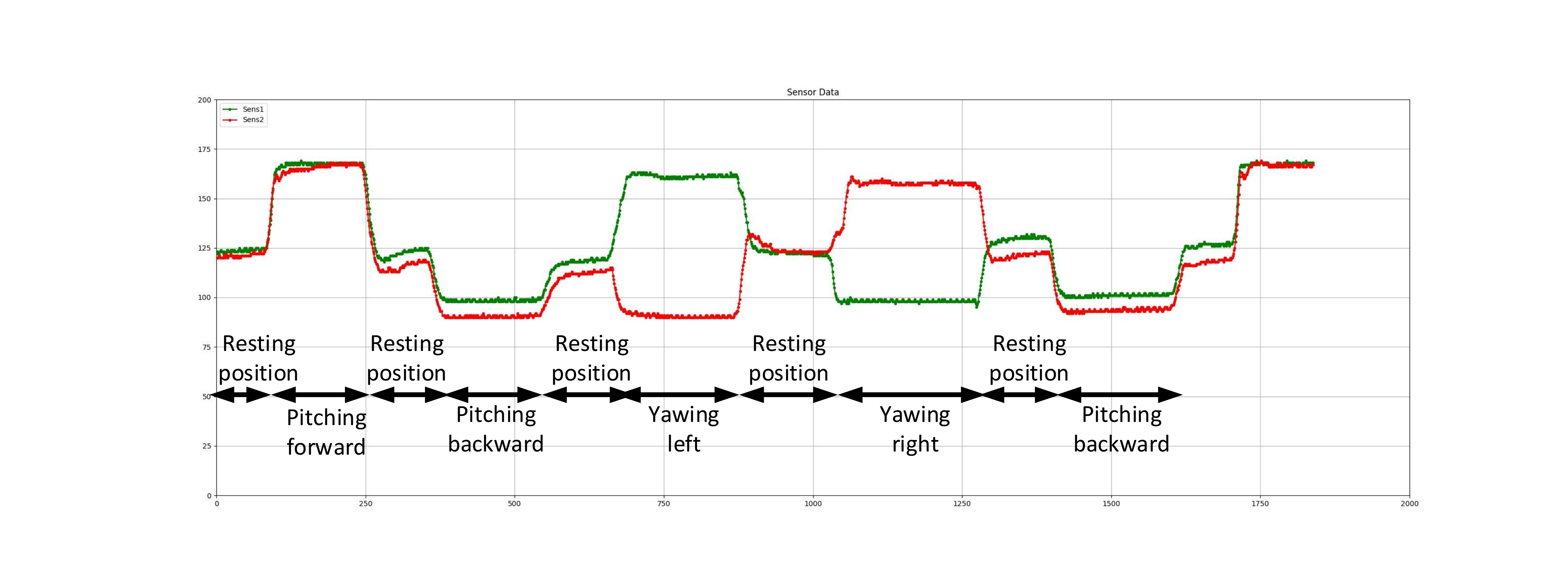}
  \caption{The signal changes during the pitching and yawing movement.}
  \label{signal}
  \end{figure} 
\section{Methods}
\label{sec:methods}
The proposed algorithm for controlling a cursor on a screen is based on the distance between the IR sensor and the user's head (chin). In resting position, two sensors approximately receive the same amount
of IR signal due to symmetry. Tilting down the user's head brings both sensors
closer to the surface of reflection, resulting in a stronger signal. Likewise,
tilting up the head weakens the reflection received by both sensors due to the
longer distance.
When the head turns to the right, the left sensor sees a
weak reflection from the left side of the chin, but the right sensor sees little
change in the reflection since the chin is still above the receiver. Thus, a
weak signal on the left sensor and a strong signal on the right sensor indicates
a cursor movement to right.  The same is applied to yawing to the left. \FIG{signal} shows the signal values for different movements of pitching
up and down and turning to the left and right. \\

\indent The input data is filtered using moving average window method in which
the input data is the average of last 15 data samples from the sensors.
Averaging eliminates the ambient noise and helps the cursor to move smoothly on
the screen, preventing it from unwanted jumps. Also, due to the nature of IR
radiation, input noise level can vastly change. For example, the IR level in
the day or under the sunlight is significantly higher than indoor spaces. To address this issue, we used dynamic thresholds to eliminate the problem of noise level changes.
The signals go through a series of simple processing to
map the head movement to cursor movement. Two different modes of joystick and
direct mapping are implemented.
\subsection{Joystick Mode}\label{AA}
\begin{figure}
\begin{subfigure}[\quad]
\centering
\includegraphics[trim=0 0 0 0,width=40mm]{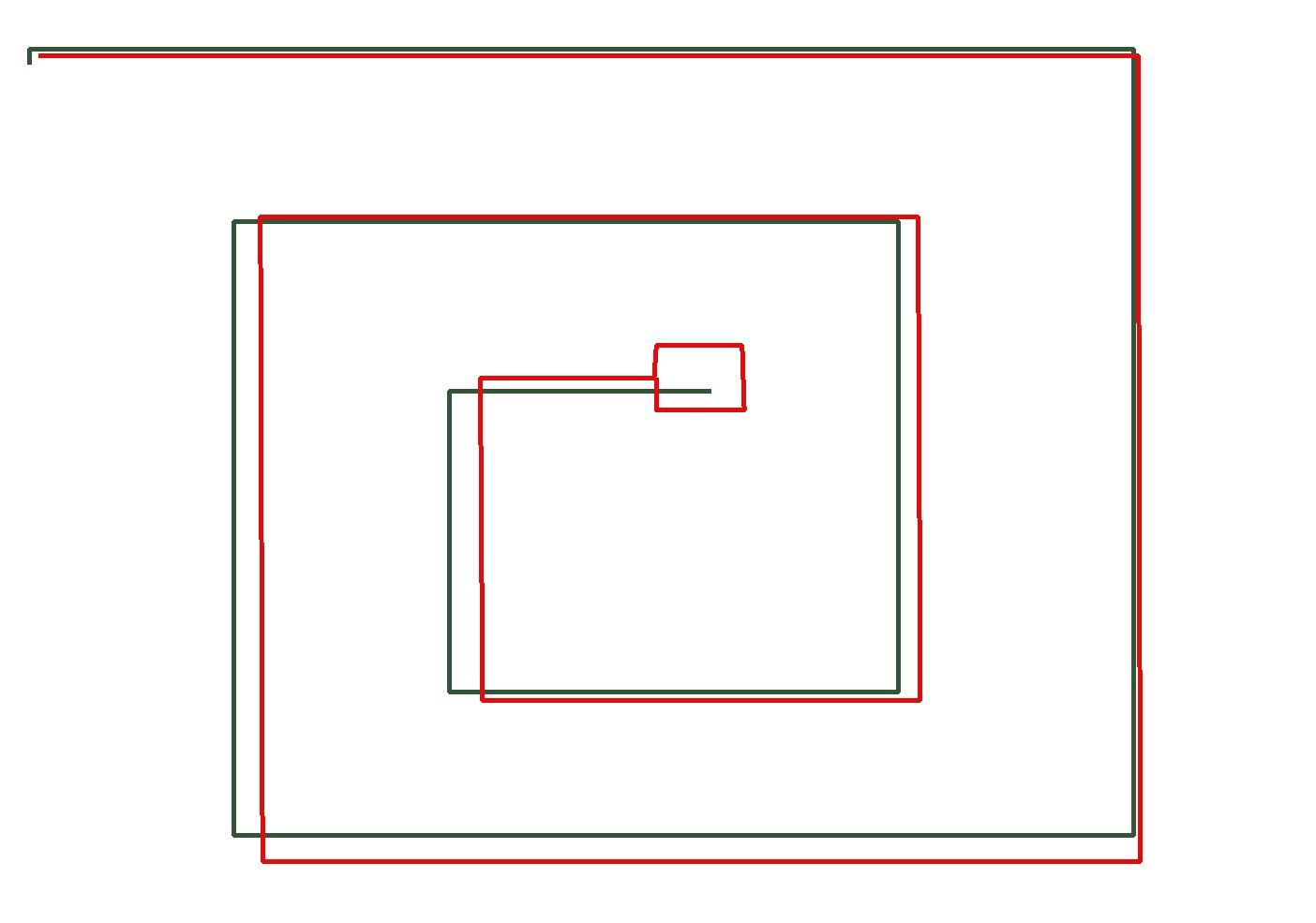}
\end{subfigure}
\begin{subfigure}[\quad]
\centering
\includegraphics[trim=0 0 0 0,width=40mm]{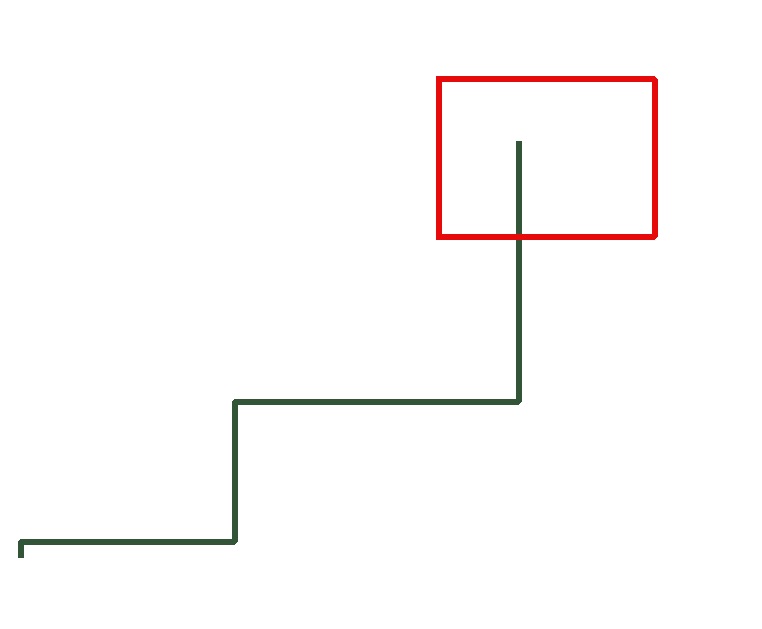}
\end{subfigure}
\begin{subfigure}[\quad]
\centering
\includegraphics[trim=0 0 0 0,width=40mm]{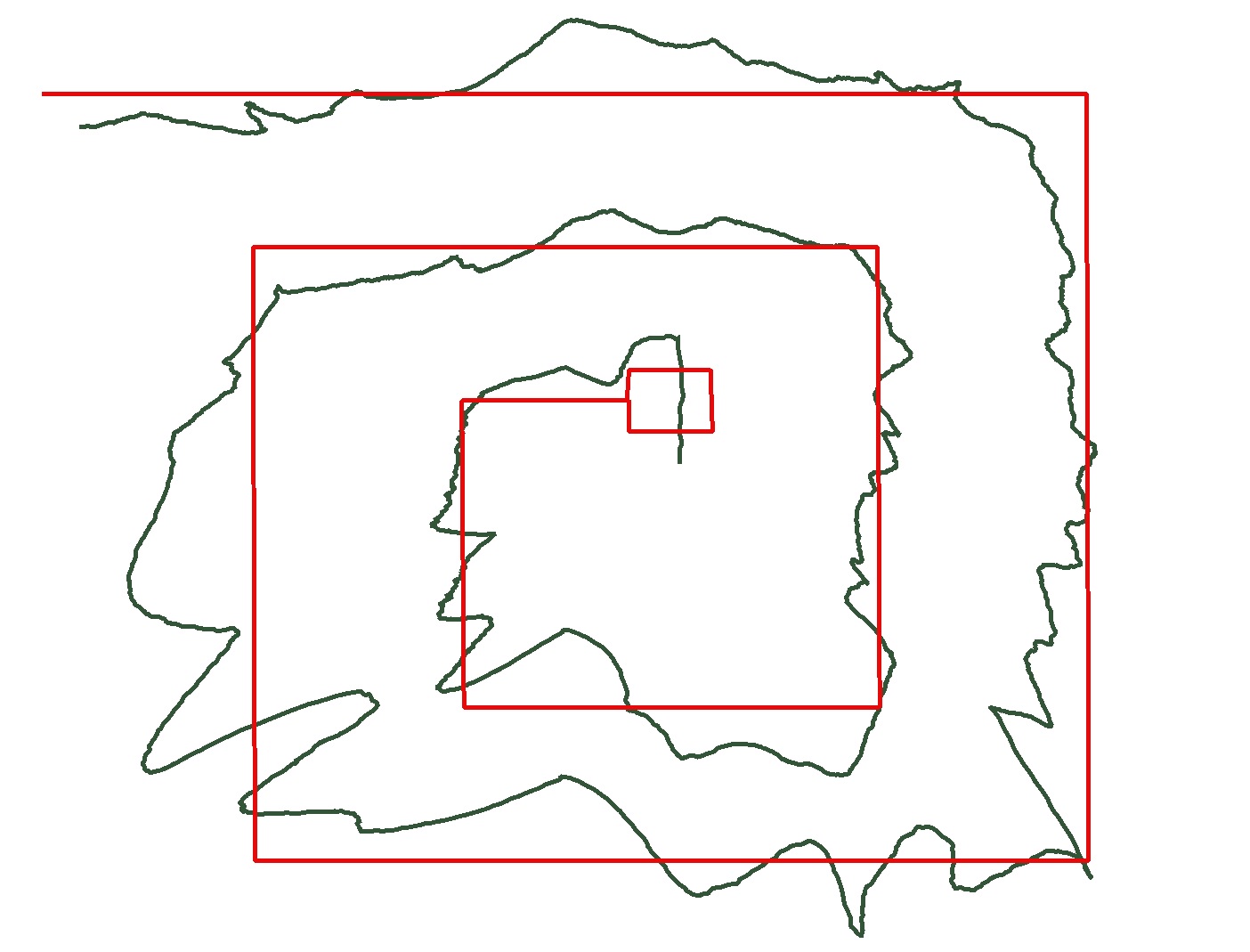}
\end{subfigure}
\begin{subfigure}[\quad]
\centering
\includegraphics[trim=0 0 0 0,width=40mm]{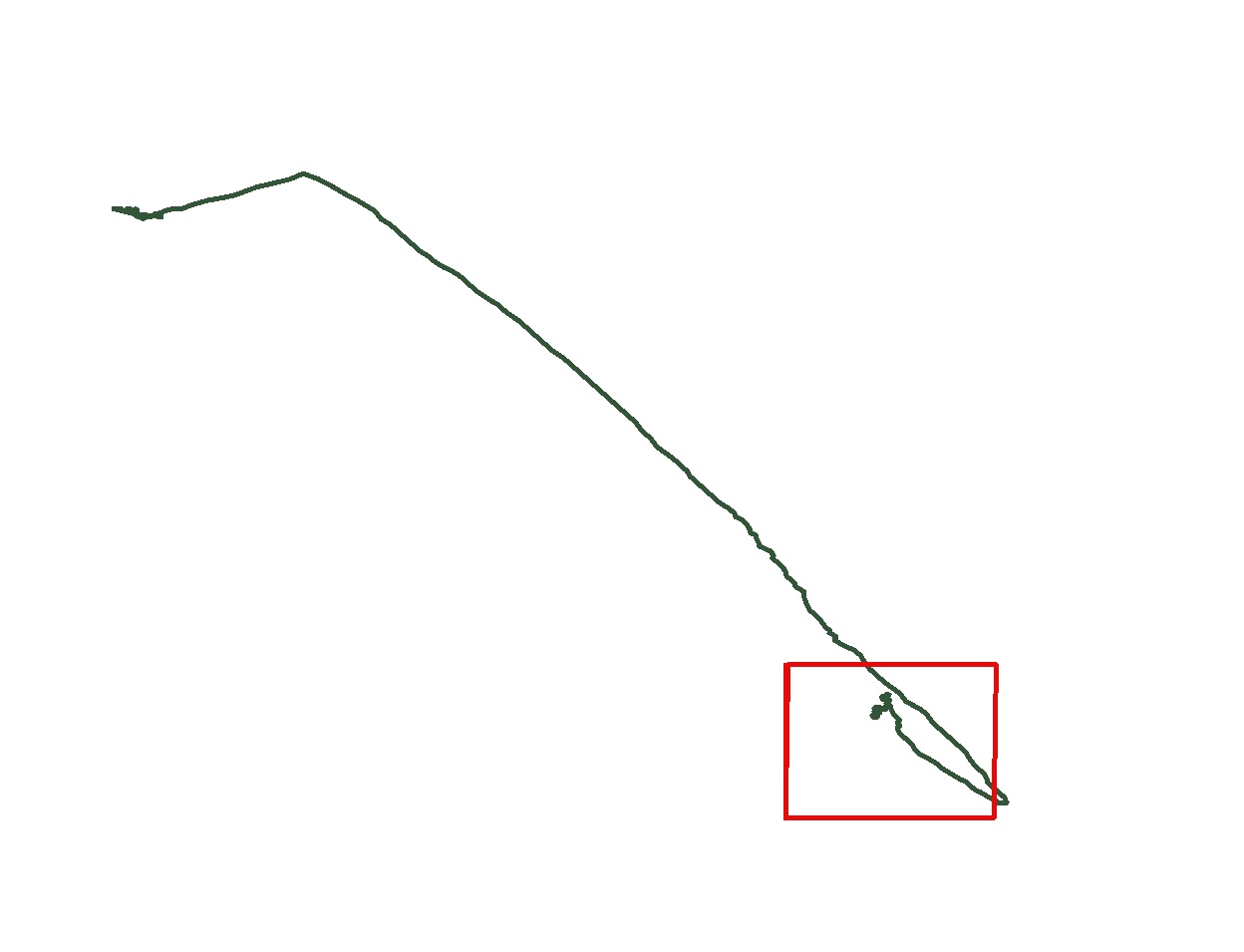}
\end{subfigure}

\caption{The testing platforms, (a) User tries to follow a predefined path in the joystick mode. (b) User tries to guide the cursor into a target in the joystick mode. (c) Following the path in direct mapping mode. (d) Guiding the cursor to the target in direct mapping mode. }
\label{pattern1}
\end{figure}

 In joystick mode, the cursor
moves only horizontally and vertically with a predefined constant speed. The data is used to recognizes the direction of the movement. For each sensor, a lower and an upper threshold
are adopted during calibration time at the initial 0.1 second of performance
where user maintains the resting head position. These thresholds are denoted
by $\id{th}^{(1)}_{\id{lower}}$, $\id{th}^{(1)}_{\id{upper}}$,
$\id{th}^{(2)}_{\id{lower}}$, $\id{th}^{(2)}_{\id{upper}}$, where the superscripts 1 and 2 represent the values for sensors 1 and 2. Algorithm
\ref{calib_joy} shows the pseudo code for adjusting the upper and lower
thresholds for both the sensors. The average of the raw data from the sensors
in the calibration mode is denoted by $\id{S1}_{\id{initial}}$ and
$\id{S2}_{\id{initial}}$.
\begin{algorithm}[t]
\caption{Joystick calibration: Adjusting the thresholds. }
\label{calib_joy}
$\bullet$ \textbf{Collect the samples during the calibration phase. Denote
the average of the samples by } $\id{S1}_{\id{initial}}$ \textbf{and}
$\id{S2}_{\id{init}}$\par
$\bullet$ \textbf{Set}  $\quad \id{th}^1_{\id{lower}}\gets\id{S1}_{\id{init}}-5$ \par
$\quad \quad \quad \id{th}^2_{\id{lower}}\gets\id{S2}_{\id{init}}-5$ \par
	$\quad \quad \quad \id{th}^1_{\id{upper}}\gets\id{S1}_{\id{init}}+10$ \par
		$\quad \quad \quad \id{th}^2_{\id{upper}}\gets\id{S2}_{\id{init}}+10$ \par
\end{algorithm}
 This calibration helps us
adjust the thresholds based on the ambient light and IR noise in different
situations. Once the thresholds are learned, the cursor moves based on the algorithm described in algorithm \ref{joy}.\\
We tried to follow a
pre-drawn pattern on the screen and reach a target as depicted in
\FIG{pattern1}. The path traveled by the cursor which is controlled by the user is drawn
in a black line. In \FIG{pattern1} (a), the user is following the path in red,
while in \FIG{pattern1} (b), the user tries to move the cursor inside the target
depicted with a square. To qualify the test scenario as successful, the cursor should stay inside the target region  for two seconds.
\begin{algorithm}[t]
\caption{Joystick mode. }
\label{joy}
$\bullet$ \textbf{ If}\quad \quad\big($\id{S1}(t)>\id{th}^{(1)}_{\id{upper}}$
\quad and \quad $\id{S2}(t)>\id{th}^{(2)}_{\id{upper}}$\big)\par
$\quad \quad Y\gets Y-1; $~~\small{Y is the vertical cursor coordinate. Move the cursor one pixel downward.}\par
$\bullet$ \textbf{ If}\quad \quad\big($\id{S1}(t)<\id{th}^{(1)}_{\id{lower}}$
\quad and \quad $\id{S2}(t)<\id{th}^{(2)}_{\id{lower}}$\big)\par
$\quad \quad Y\gets Y+1; $~~\small{ Move the cursor one pixel downward.}\par
$\bullet$ \textbf{ If}\quad \quad\big($\id{S1}(t)>\id{th}^{(1)}_{\id{upper}}$
\quad and \quad $\id{S2}(t)<\id{th}^{(2)}_{\id{lower}}$\big)\par
$\quad \quad X\gets X-1; $~~\small{X is the horizontal cursor coordinate. Move the cursor one pixel to the left.}\par
$\bullet$ \textbf{ If}\quad \quad\big($\id{S1}(t)<\id{th}^{(1)}_{\id{lower}}$
\quad and \quad $\id{S2}(t)>\id{th}^{(2)}_{\id{upper}}$\big)\par
$\quad \quad X\gets X+1; $~~\small{ Move the cursor one pixel to the right.}\par
\end{algorithm}

\subsection{Direct-Mapping Mode}
In direct-mapping mode, the absolute value of the sensor data is directly mapped to the cursor
location on the GUI.
If the user tilts the head up to the maximum extend, the
cursor touches the upper border. Similarly, the extreme positions of the user's head in tilting down and yawing to the right and left are mapped to the borders of the GUI screen. Thus, we need to define the range of head
movement so that we can map it to the pointer location. A set of training movements
should be conducted in the calibration phase at the first 4 seconds of the performance. The
user is required to maximally move his head to up, down, left, and right at startup.
During this time, sensors data is being constantly checked to update the
thresholds. In this mode, four thresholds are adopted for each sensor.
$\id{Th}^{(1)}_U$, $\id{Th}^{(1)}_D$, $\id{Th}^{(1)}_L$,  and
$\id{Th}^{(1)}_R$ are specifically adopted for sensor1 in moving up, down,
left, and right directions respectively. Likewise $\id{Th}^{(2)}_U$,
$\id{Th}^{(2)}_D$, $\id{Th}^{(2)}_L$,  and $\id{Th}^{(2)}_R$ are learned for sensor 2. Algorithm \ref{calib_direct} shows the pseudo code for learning the thresholds in direct mapping calibration phase.
\begin{algorithm}[t]
\caption{Direct mapping mode, calibration phase. }
\label{calib_direct}
 $\bullet$ \textbf{While in Calibration do:}\par
 $\bullet$ \textbf{ If}\quad \quad\big($\id{S1}(t)<\id{Th}^{(1)}_{U}$ \quad
 and \quad $\id{S2}(t)<\id{Th}^{(2)}_{U}$\big)\par
 $\quad \quad$ \textbf{Update}$\quad \id{Th}^{1}_U\gets\id{S1}(t)$\par
 $\quad \quad \quad \quad \quad \quad \id{Th}^{2}_U\gets\id{S2}(t)$\par
 $\bullet$ \textbf{ If}\quad \quad\big($\id{S1}(t)>\id{Th}^{(1)}_{D}$ \quad
 and \quad $\id{S2}(t)>\id{Th}^{(2)}_{D}$\big)\par
 $\quad \quad$ \textbf{Update}$\quad \id{Th}^{1}_D\gets\id{S1}(t)$\par
 $\quad \quad \quad \quad \quad \quad \id{Th}^{2}_D\gets\id{S2}(t)$\par
 $\bullet$ \textbf{ If}\quad \quad\big($\id{S1}(t)>\id{Th}^{(1)}_{L}$ \quad
 and \quad $\id{S2}(t)<\id{Th}^{(2)}_{L}$\big)\par
 $\quad \quad$ \textbf{Update}$\quad \id{Th}^{1}_L\gets\id{S1}(t)$\par
 $\quad \quad \quad \quad \quad \quad \id{Th}^{2}_L\gets\id{S2}(t)$\par
 $\bullet$ \textbf{ If}\quad \quad\big($\id{S1}(t)<\id{Th}^{(1)}_{R}$ \quad
 and \quad $\id{S2}(t)>\id{Th}^{(2)}_{R}$\big)\par
 $\quad \quad$ \textbf{Update}$\quad \id{Th}^{1}_R\gets\id{S1}(t)$\par
 $\quad \quad \quad \quad \quad \quad \id{Th}^{2}_R\gets\id{S2}(t)$\par
\end{algorithm}

 After 4 seconds, thresholds are fixed and put in to the equations. There are three equations, one governs the changes in Y coordinate of the cursor and the other two are used to determine X coordinate. For Y we have:
\begin{equation}\label{Y}
\text{Y} = ( Yaxis \times \frac{avg(S1(t),S2(t))}{\big( avg(Th^{(1)}_U,Th^{(2)}_U) - avg(Th^{(1)}_D,Th^{(2)}_D)  \big)}
\end{equation}


where \id{Yaxis} is the number of pixels in $Y$ direction and $\text{avg}()$ finds the average its inputs.
For X coordinate, we had to use two equations. One controls the position in the left half of the screen and the other one controls it in the right half of the screen.
\begin{equation}
	\id{X} = \left(1-
	\frac{\left(\id{S1}(t)-\id{S2}(t)\right)}{\left(\id{Th}^{(1)}_L -
	\id{Th}^{(2)}_L\right)} \right)\times \frac{\id{Xaxis}}{2}\\
\end{equation}
\begin{equation}
\text{X} = \frac{(S2-S1)} {( Th^{(2)}_R-Th^{(1)}_R )}\times(\frac{Xaxis}{2}) + \frac{Xaxis}{2}
\end{equation}

\subsection{Performance Metrics}
According to Fitt's law, there is an inherent trade-off between the speed and accuracy. We
will resort to the performance metrics mostly used in the state-of-art
literature to quantify the performance. These metrics including, the index of difficulty, path efficiency, throughput, and overshoot are defined as follows:\\
\textbf{Index of Difficulty (ID)}: ID is defined as
\begin{equation}
\id{ID}=\frac{\id{D}}{\id{W}},
\end{equation}
where $\id{D}$ is the distance between the original location of the cursor
and the center of the target, and \id{W} is the width of the target.\\
\textbf{Path Efficiency (PE)}:
Path efficiency is a measure of straightness of the path the cursor is
traveling and is defined as the ratio of the total distance between the original location of the cursor and target to the path length traveled by the cursor to :
\begin{equation}
\id{PE}=\frac{\id{D}}{\id{P}},
\end{equation}
where \id{P} is the length of the path traveled by the cursor.\\
\textbf{Throughput (TP)}:
Throughput is a metric for measuring how fast the cursor is moving and is defined as:
\begin{equation}\label{TP}
\id{TP}=\frac{\id{ID}}{\id{MT}},
\end{equation}
where \id{MT} is the time it takes for the user to move the cursor to the
target zone.

\section{Results}
\label{sec:results}
\begin{figure}
\centering
 \includegraphics[trim=0 0 0 0,width=\columnwidth]{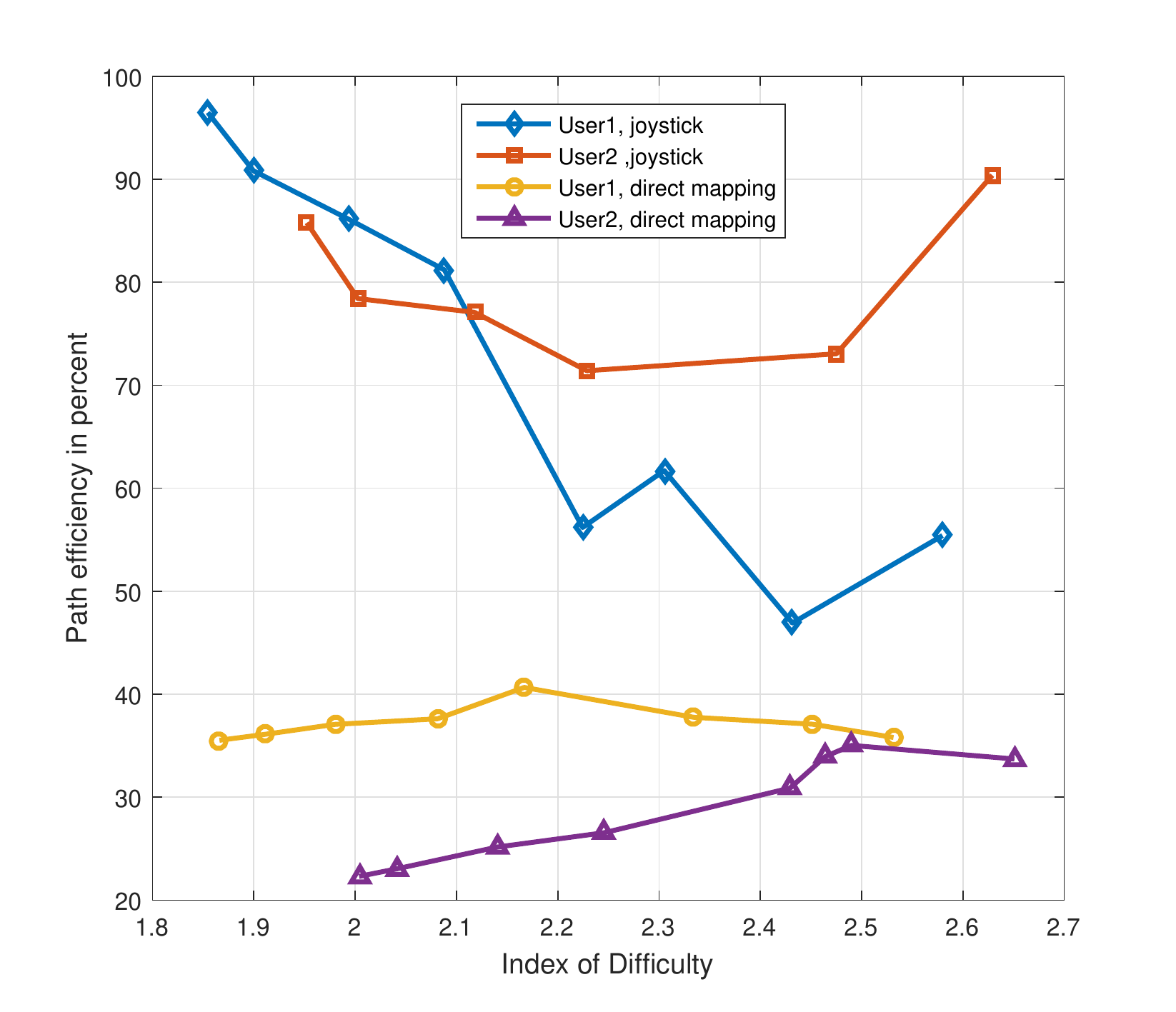}
\caption{Path efficiency with respect to index of difficulty for two modes of joystick and direct mapping.}
\label{PE}
\end{figure}
\begin{figure}
\centering
 \includegraphics[trim=0 0 0 0,width=\columnwidth]{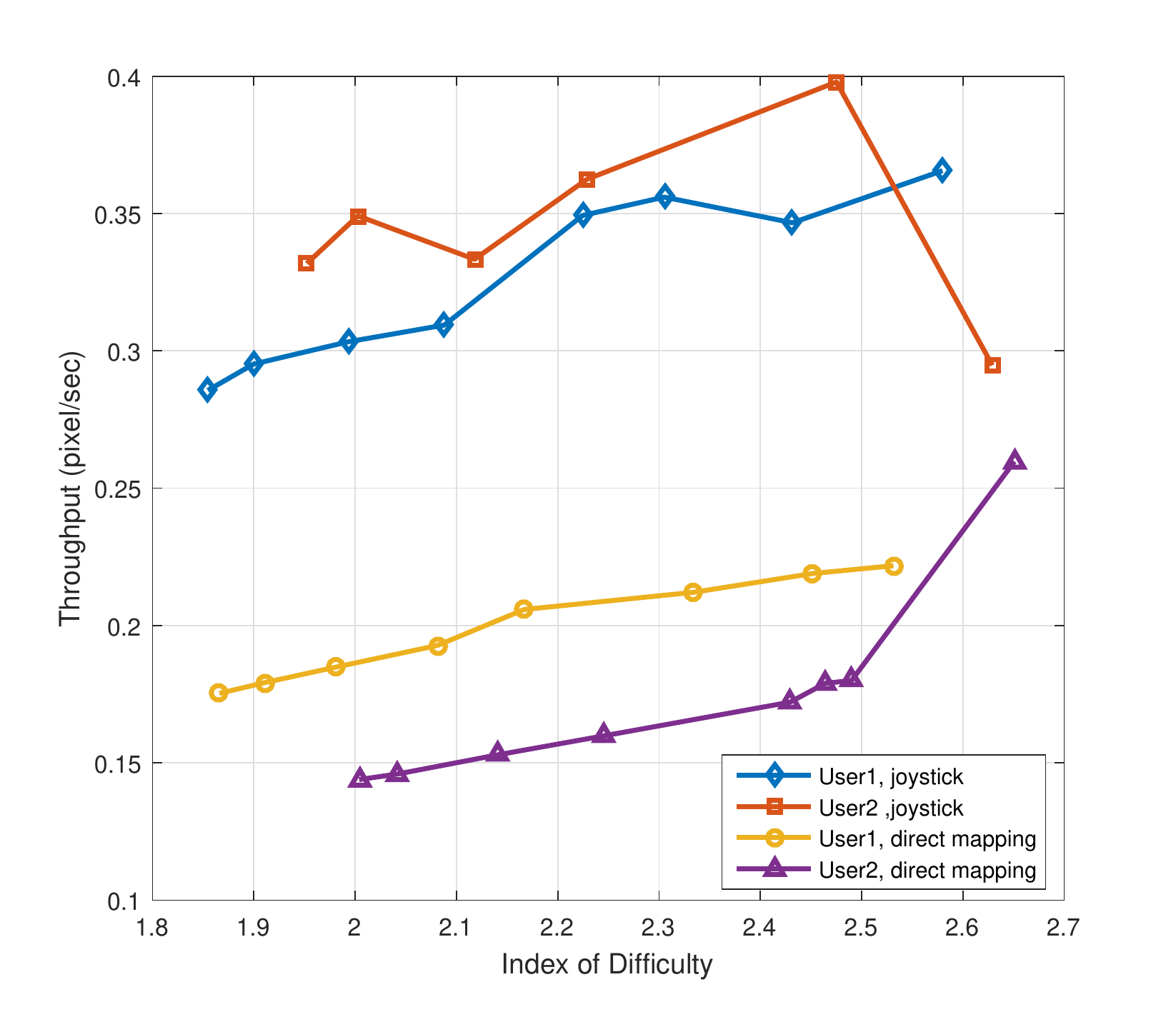}
\caption{Throughput with respect to index of difficulty for two modes of joystick and direct mapping.}
\label{PE}
\end{figure}

Two different users volunteered to test the implemented system in the two
modes. The task requires the users to move the cursor into a target. The
location of the target on the screen and its width is randomly chosen, and
the task is repeated. In fact, the index of difficult is a function of the
location and the size of the target. The GUI automatically keeps the records
of the moving time (\id{MT}) and the path length $P$. The number of
trials for both joystick mode and direct mapping mode is 50 each.

\FIG{PE} shows the path efficiency of the users for both the joystick and
direct mapping modes, while \FIG{TP} demonstrates the throughput. It is
observed in the path efficiency and throughput are significantly better
in the joystick mode compared to the direct mapping mode. This is observable
from \FIG{pattern1} where the user has a better controllability in joystick
mode compared to direct mapping.

\section{Conclusions}
\label{sec:concl}

We have presented a new wireless wearable device for controlling a pointing
device by head movements.  It uses conventional IR sensors to sense head
movements, and a commodity MCU to map the sensor data to either relative
movement (joystick mode) or absolute location (direct mapping mode) on the
screen.  While the principle of operation is relatively straightforward, in
practice, the system must address several challenges, including optical noises
from different sources, the wide range of head movement on different users,
accuracy, and power consumption. We solve these problems by using adaptive
calibration for sensors, algorithm for coordinate generation, and PWM control.
Several directions for future development remain. We plan to continue
improving accuracy by better filtering of input signals and more tuning of threshold. For now, the head needs to have a considerable rotation to detect movement that causes a feeling of tiredness. By increasing the accuracy, we can reduce the range of head movement, resulting in a better and more comfortable user experience.
Adding a clicking procedure based on a specific pattern of movement is also considered.
For technical aspects, we aim to improve power efficiency and enable plug-and-play by performing
calculations on the device and communicating with the host computer directly
using HID profile. It will enable the user not only to control
a general-purpose PC but also embedded devices that can assist the user with a
variety of tasks, including navigating a wheel chair, a robot arm, and many
other assistive technologies to be invented.


\section*{Acknowledgment}

The authors would like to thank
 Cho-An Lin and Hsun-Fa Zhou
at
National Tsing Hua
University for their help with the BLE communication between the device and
the host.

\bibliographystyle{IEEEtran}
\bibliography{headmouse}
\end{document}